\documentclass[10pt,aps,prb,twocolumn,amsmath,amssymb,superscriptaddress,nobibnotes]{revtex4-1}
\usepackage{amsmath}
\usepackage{graphicx}
\usepackage{color}
\usepackage{multirow}
\usepackage{caption}
\usepackage[caption = false]{subfig}
\usepackage{array}
\usepackage{threeparttable}
\usepackage[version=4]{mhchem}
\usepackage{mathtools}
\usepackage{dsfont}
\usepackage{siunitx}
\usepackage[skip=7pt plus1pt, indent=0pt]{parskip}
\usepackage{centernot}
\usepackage{tikz}
\usepackage{hyperref}
\usepackage{xurl}
\usepackage{bm}
\hypersetup{breaklinks=true}
\usepackage{appendix}
\usepackage{float}

\usepackage[capitalise]{cleveref}
\tikzset{font={\fontsize{11pt}{12}\selectfont}}
\usetikzlibrary{shapes,arrows}
\usepackage{lipsum}
\usepackage[normalem]{ulem}
\makeatletter

\renewcommand*{\p@subsection}{}

\renewcommand*{\p@subsubsection}{}
\makeatother

\usepackage{hyperref}
\hypersetup{linkcolor=blue, citecolor=blue, urlcolor=blue}
\usepackage{xcolor}

\newcommand{\ket}[1]{|#1\rangle}
\newcommand{\bra}[1]{\langle#1|}

\usepackage{acro}

\newcommand{\Harvard}{\affiliation{Department of Chemistry and Chemical Biology, Harvard University, Cambridge, MA, USA}}

\begin{document}
\title{Regularized Perturbation Theory for {\it Ab initio} Solids}
\author{Meng-Fu Chen}
\author{Jinghong Zhang}
\author{Hieu Q. Dinh}
\author{Adam Rettig}
\author{Joonho Lee}
\email{joonholee@g.harvard.edu}
\Harvard

\begin{abstract}
\noindent Second-order M{\o}ller-Plesset perturbation theory (MP2) for \textit{ab initio} simulations of solids is often limited by divergence or over-correlation issues, particularly in metallic, narrow-gap, and dispersion-stabilized systems. We develop and assess three regularized second-order perturbation theories: $\kappa$-MP2, $\sigma$-MP2, and the size-consistent Brillouin--Wigner approach (BW-s2), across metals, semiconductors, molecular crystals, and rare gas solids. BW-s2 achieves high accuracy for cohesive energies, lattice constants, and bulk moduli in metals, semiconductors, and molecular crystals, rivaling or surpassing coupled-cluster with singles and doubles at lower cost. In rare gas solids, where MP2 already underbinds, $\kappa$-MP2 does not make the results much worse while BW-s2 struggles. These results illustrate both the potential and the limitations of regularized perturbation theory for efficient and accurate solid-state simulations. While broader testing is warranted, BW-s2($\alpha$ = 2) appears particularly promising, with possible advantages over modern random-phase approximations and coupled-cluster theory.
\end{abstract}

\maketitle

{\it Introduction.}
\textit{Ab initio} simulations of materials using quantum chemistry methods have become an active research area in the past few years.~\cite{robinson_condensed-phase_2025} Kohn-Sham density functional theory (KS-DFT) has long been the workhorse method for quantum chemistry and materials science due to its favorable $\mathcal{O}(N^3)$ scaling with respect to system size ($N$) and good accuracy for a wide variety of problems.\cite{hasnip_density_2014, mardirossian_thirty_2017, teale_dft_2022}  KS-DFT suffers from fundamental flaws such as delocalization error,\cite{cohen_insights_2008} self-interaction error,\cite{huang_advances_2008} and difficulties in describing long-range dispersion.\cite{grimme_dispersion-corrected_2016} These translate into well-known problems in solid state systems such as the underestimation of band gaps \cite{perdew_density_1985} and disagreements on the chemisorption energies of small molecules on metallic surfaces.\cite{lazic_density_2010, janthon_adding_2017} However, it is worth noting that incorporating exact exchange as in hybrid functionals often provides improved results.\cite{garza_predicting_2016, stroppa_shortcomings_2008, lee_faster_2022}

In contrast, wavefunction methods are known to give ``the right answer for the right reasons", \cite{teale_dft_2022} but at the expense of much higher computational cost.
For instance, coupled-cluster based methods like coupled-cluster singles and doubles (CCSD) and perturbative triples (CCSD(T)) are well known to be accurate for insulating solid state systems \cite{booth_towards_2013, yang_ab_2014, gruber_applying_2018} and has been recently benchmarked for simple metallic systems, \cite{neufeld_ground-state_2022, PhysRevLett.131.186402} but the high computational cost ($\mathcal{O}(N^6)$ for CCSD, $\mathcal{O}(N^7)$ for CCSD(T)) and the storage cost of wavefunction amplitudes makes them challenging to use for realistic simulations. 
Therefore, there has been considerable effort to develop lower-scaling wavefunction methods either by local correlation methods \cite{maschio_local_2011, muller_local_2011, usvyat_linear-scaling_2013, wang_cluster--molecule_2022, ye_periodic_2024} or by developing simpler methods. \cite{del_ben_electron_2013, schafer_quartic_2017, haritan_efficient_2025}   

Currently, the state-of-the-art $\mathcal{O}(N^4)$ method for solid state systems is the random-phase approximation, which has its roots in the uniform electron gas model.\cite{bohm_collective_1951, pines_collective_1952, bohm_collective_1953} 
Note that here we only consider the direct part in the electron repulsion integrals, also known as direct random phase approximation (dRPA) in the literature. 
Various approaches have been proposed to improve the accuracy of dRPA, such as second-order screened exchange (SOSEX) \cite{gruneis_making_2009}  ($\mathcal O(N^5)$) that removes self-correlation error and renormalized singles correlation energy contributions (rSE).~\cite{ren_beyond_2011, klimes_singles_2015} While other options are available, \cite{harl_accurate_2009, kaltak_low_2014, yeh_low-scaling_2023} our implementation follows the resolution-of-identity dRPA energy evaluation framework, originally proposed for molecules \cite{eshuis_fast_2010} and later extended to solids.\cite{grundei_random_2017, stein_massively_2024}  

\vspace*{-\parskip}\vspace*{10pt}    
A natural candidate for a useful $\mathcal{O}(N^5)$ method is second-order Møller–Plesset perturbation theory (MP2) where, unlike dRPA, no self-correlation error exists. For periodic systems, the MP2 correlation energy per unit cell is:\cite{marsman_second-order_2009}
\begin{equation}\label{eq:emp2}
E_{c}^{\mathrm{MP2}}
=
\frac{1}{4 N_k} \sum_{ijab}\sum_{\mathbf{k}_1\mathbf{k}_2\mathbf{q}}
\epsilon_{i_{\mathbf k_1}j_{\mathbf k_2+\mathbf q}}^{a_{\mathbf k_1+\mathbf q}b_{\mathbf k_2}},
\end{equation}
where
\begin{equation}
\epsilon_{i_{\mathbf k_1}j_{\mathbf k_2+\mathbf q}}^{a_{\mathbf k_1+\mathbf q}b_{\mathbf k_2}} =
\frac{|\langle i_{\mathbf{k}_1}j_{\mathbf{k}_2+\mathbf{q}}||a_{\mathbf{k}_1+\mathbf{q}}b_{\mathbf{k}_2}\rangle|^2}
{\Delta_{i_{\mathbf k_1}j_{\mathbf k_2+\mathbf q}}^{a_{\mathbf k_1+\mathbf q}b_{\mathbf k_2}}},
\end{equation}
with
\begin{equation}
\Delta_{i_{\mathbf k_1}j_{\mathbf k_2+\mathbf q}}^{a_{\mathbf k_1+\mathbf q}b_{\mathbf k_2}}
=
\varepsilon_{i_{\mathbf{k}_1}} + \varepsilon_{j_{\mathbf{k}_2+\mathbf{q}}} - \varepsilon_{a_{\mathbf{k}_1+\mathbf{q}}} - \varepsilon_{b_{\mathbf{k}_2}},
\end{equation}
where $\{\varepsilon_{i_\mathbf{k}}\}$ is the $i$-th occupied band energy at $\mathbf{k}$ and $\{\varepsilon_{a_\mathbf{k}}\}$ is the $a$-th unoccupied band energy at $\mathbf{k}$. We use $\{i,j\}$ and $\{a,b\}$ to denote occupied and unoccupied bands, respectively.  Evaluation of \cref{eq:emp2} scales as $\mathcal{O}(N_k^3N^5)$ due to integral transformation or the assembly of density fitting or Cholesky factors. \cite{marsman_second-order_2009, del_ben_second-order_2012}
The performance of MP2 for various solid state systems has been comprehensively benchmarked in the past, including typical insulators and semiconductors, \cite{marsman_second-order_2009, gruneis_second-order_2010, lange_improving_2021} molecular crystals, \cite{bintrim_integral-direct_2022, liang_can_2023} and surface adsorption problems.~\cite{keller_regularized_2022, ye_adsorption_2024} 
An inherent limitation of MP2 is its divergence when the energy denominator goes to zero, which severely limits its utility for metallic systems and narrow-gap materials. Furthermore, even in the absence of divergence, MP2 is notorious for overestimating non-covalent interaction energies.\cite{villard_plane_2023}  

One way to remedy these is by regularization that removes the divergence and dampens away unphysical correlation contributions. 
There have been various regularization schemes proposed to this day.\cite{assfeld_degeneracy-corrected_1995, surjan_damping_1996, szabados_near-degeneracy_2001, stuck_regularized_2013, shee_regularized_2021, coveney_regularized_2023, sawicki_analysis_2025} 
In this work, we focus on the orbital-energy-dependent regularizers proposed by Lee and Head-Gordon,~\cite{lee_regularized_2018} namely the $\kappa$ and $\sigma$ regularizers. $\kappa$-MP2 correlation energy per cell reads, with some regularization constant $\kappa$,
\begin{equation}\label{eq:kappa}
E_{c}^{\kappa\textrm{-MP2}}
=
\frac{1}{4N_k} \sum_{ijab}\sum_{\mathbf{k}_1\mathbf{k}_2\mathbf{q}}
\epsilon_{i_{\mathbf k_1}j_{\mathbf k_2+\mathbf q}}^{a_{\mathbf k_1+\mathbf q}b_{\mathbf k_2}}
\left(1 - e^{-\kappa\Delta_{i_{\mathbf k_1}j_{\mathbf k_2+\mathbf q}}^{a_{\mathbf k_1+\mathbf q}b_{\mathbf k_2}}}\right)^2
\end{equation}
and $\sigma$-{MP2} correlation energy per cell is, with some regularization constant $\sigma$,
\begin{equation}\label{eq:sigma}
E_{c}^{\sigma\textrm{-MP2}}
= 
\frac{1}{4N_k} \sum_{ijab}\sum_{\mathbf{k}_1\mathbf{k}_2\mathbf{q}}
\epsilon_{i_{\mathbf k_1}j_{\mathbf k_2+\mathbf q}}^{a_{\mathbf k_1+\mathbf q}b_{\mathbf k_2}}
\left(1 - e^{-\sigma\Delta_{i_{\mathbf k_1}j_{\mathbf k_2+\mathbf q}}^{a_{\mathbf k_1+\mathbf q}b_{\mathbf k_2}}}\right).
\end{equation}

Both of these regularized MP2 methods have a well-defined correlation energy even when $\Delta \rightarrow 0$, making them potentially useful methods for treating metallic systems.\cite{keller_regularized_2022} Beyond correcting the divergence of MP2, in molecular quantum chemistry, it has also been shown that orbital energy-dependent regularized MP2 better describes dispersion-dominated intermolecular interactions. \cite{shee_regularized_2021, bertels_third-order_2019, loipersberger_exploring_2021} On the other hand, for periodic systems, recent works have reported mixed performances for $\kappa$-MP2 in various model solid-state systems, \cite{keller_regularized_2022, coveney_regularized_2023, sawicki_analysis_2025} and it is yet unclear if these are broadly useful methods for studying {\it ab initio} solids. 

Apart from the aforementioned regularization strategy, another viable way to mitigate the divergent correlation energy problem is to use the framework of Brillouin-Wigner perturbation theory~\cite{brillouin_problemes_1932, wigner_modification_1997}  that nominally lacks size consistency and extensivity.~\cite{march_many-body_1995} 
By repartitioning the Hamiltonian in Brilliouin-Wigner perturbation theory, a size-consistent and size-extensive second-order method, called BW-s2($\alpha$), was proposed by Carter-Fenk and Head-Gordon. \cite{carter-fenk_repartitioned_2023, carter-fenk_optimizing_2023,dittmer_repartitioning_2025}
A complete derivation of BW-s2($\alpha$) for periodic systems will be presented in \cref{app:A}, but in short, within this framework, we construct a regularizer $\mathbf{W}_{\mathbf k}$ of the form (suppressing $\mathbf{k}$ indices):
\begin{align}
    W_{ij} &= \frac{1}{8} \sum_{kab} \left[ t_{ik}^{ab}\langle jk||ab \rangle + t_{jk}^{ab}\langle ik||ab \rangle \right]
\end{align}
where $\mathbf{t}$ are the first-order MP amplitudes using $\tilde{\epsilon}$ (see \cref{eq:dressed}). 
$\mathbf{W}_\mathbf k$ is added to the occupied-occupied block of the Fock matrix ($\mathbf{F}_{{\mathbf{k}},\mathrm{oo}}$), leading to a self-consistent equation, with some regularization parameter $\alpha$,
\begin{equation}\label{eq:dressed}
    \left(\mathbf{F}_{{\mathbf{k}},\mathrm{oo}} + \alpha \mathbf{W}_{\mathbf{k}}(\Tilde{\varepsilon}_{\mathbf{k},o})\right)\mathbf{U}_{\mathbf k} = \Tilde{\varepsilon}_{\mathbf{k},o}\mathbf{U}_{\mathbf k},
\end{equation}
where $\{\Tilde{\varepsilon}_{\mathbf{k},o}\}$ are modified occupied orbital energies at each $\mathbf k$-point. \cref{eq:dressed} is solved iteratively until self-consistency is reached. The BW-s2($\alpha$) correlation energy per unit cell is:
\begin{equation}
E_{c}^{\textrm{BW-s2}} = 
\frac{1}{4N_k} \sum_{ijab}
\sum_{\mathbf{k}_1\mathbf{k}_2\mathbf{q}}
\frac{|\langle \Tilde{i}_{\mathbf{k}_1}\Tilde{j}_{\mathbf{k}_2+\mathbf{q}}||a_{\mathbf{k}_1+\mathbf{q}}b_{\mathbf{k}_2}\rangle|^2}
{\Tilde{\varepsilon}_{i_{\mathbf{k}_1}} + \Tilde{\varepsilon}_{j_{\mathbf{k}_2+\mathbf{q}}} - \varepsilon_{a_{\mathbf{k}_1+\mathbf{q}}} - \varepsilon_{b_{\mathbf{k}_2}}}.
\end{equation}

Following the strategy for efficient implementation of molecular BW-s2($\alpha$) in Ref.~ \citenum{carter-fenk_repartitioned_2023}, our solid-state implementation BW-s2($\alpha$) scales the same as MP2, albeit with an extra factor of $N_\mathrm{iter}$ required to reach self-consistency. To our knowledge, no work has been done on implementing solid-state BW-s2($\alpha$) and evaluating its performance for a broad range of {\it ab initio} solids. 

We assess the accuracy of regularized perturbation theory methods on cohesive (or lattice) energies, equilibrium lattice constants, and bulk moduli of a few representative examples: (1) a metal (body-centered cubic lithium), (2) a wide-gap semiconductor (diamond), and (3) dispersion-dominated molecular crystals (benzene crystal, face-centered cubic neon).
All our calculations are carefully extrapolated to the thermodynamic and complete basis set limits. Relevant computational details are given in the \cref{app:B}.

\begin{figure*}[hbt!]
    \centering
    \includegraphics[width=\linewidth]{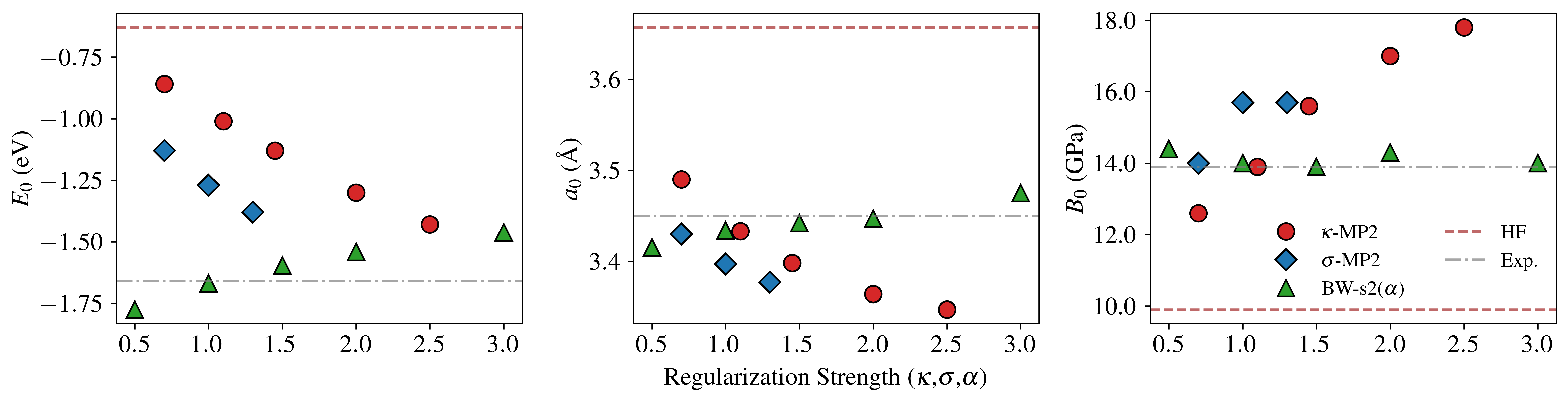}
    \caption{Cohesive energy ($E_0$), equilibrium lattice constant ($a_0$), and bulk modulus ($B_0$) of BCC Li at different regularization strengths for $\kappa$-MP2, $\sigma$-MP2, and BW-s2($\alpha$). 
    }
    \label{fig:1}
\end{figure*}

\begin{figure*}[hbt!]
    \centering
    \includegraphics[width=\linewidth]{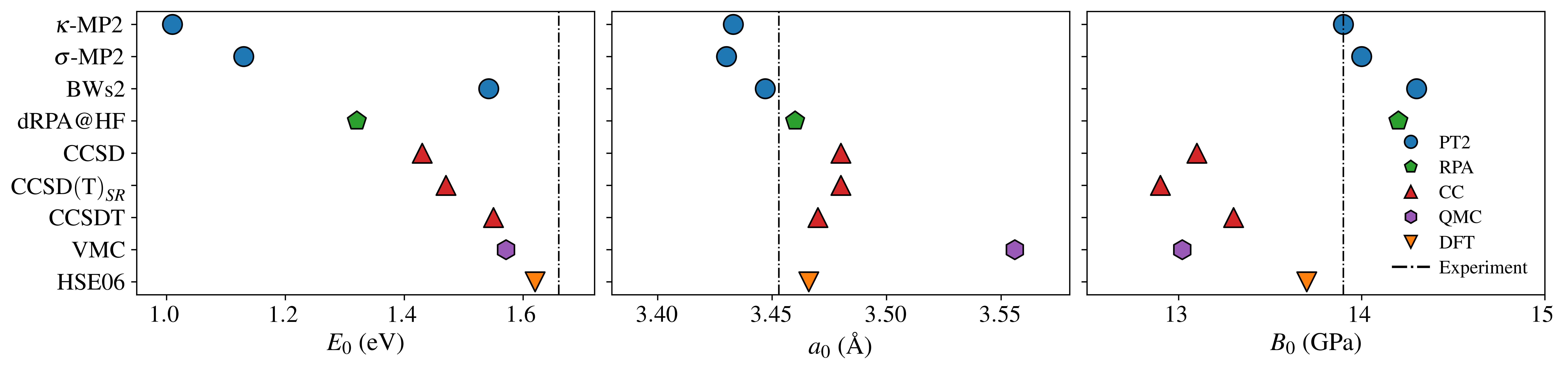}
    \caption{Cohesive energy ($E_0$), equilibrium lattice constant ($a_0$), and bulk modulus ($B_0$) of lithium calculated with different methods compared with experimental values. $\kappa = 1.1$, $\sigma = 0.7$, and $\alpha = 2.0$ are used for $\kappa$-MP2, $\sigma$-MP2, and BW-s2$(\alpha)$. CCSD result is from Ref. [\hspace{-5pt} \citenum{ye_periodic_2024}], CCSD(T)$_{SR}$ from Ref. [\hspace{-5pt} \citenum{neufeld_ground-state_2022}], and CCSDT from Ref. [\hspace{-5pt} \citenum{PhysRevLett.131.186402}]. VMC result is from Ref. [\hspace{-5pt} \citenum{PhysRevB.54.8393}]. HSE06 result is from Ref. [\hspace{-5pt} \citenum{zhang_performance_2018}]. Experimental result is from Ref. [\hspace{-5pt} \citenum{schimka_improved_2011}] and includes ZPE correction.}
    \label{fig:2}
\end{figure*}

{\it BCC lithium.}
MP2 correlation energy diverges for metals due to the vanishing denominator, and hence, it is important to understand the utility of regularized methods for metals.
A body-centered cubic (BCC) Li crystal was chosen as a representative metallic system, since reliable theoretical and experimental results exist to benchmark against.\cite{kittel_introduction_2007, zhang_performance_2018, schimka_improved_2011, PhysRevLett.131.186402, PhysRevB.54.8393, rasch_fixed-node_2015} The zero-point energy (ZPE) corrected experimental values are 1.66 eV for cohesive energy, 3.45~{\AA} for equilibrium lattice constant, and 13.9 GPa for bulk modulus. \cite{schimka_improved_2011}
In \cref{fig:1}, we examine the effect of regularization strength on cohesive energy, equilibrium lattice constant, and bulk modulus.
With sufficiently strong regularization strengths, all three methods, $\kappa$-MP2, $\sigma$-MP2, BW-s2($\alpha$), approach the HF result (0.63 eV/atom, 3.66~{\AA}, 9.9 GPa).  
We note that the bulk modulus computed by BW-s2 remains nearly constant over the range of regularization strengths we examined.

One crucial observation is that the regularization strength used in both $\kappa$- and $\sigma$-MP2 not only significantly affects the cohesive energy, but also shifts the predicted equilibrium lattice constant drastically. This is in contrast to molecular systems, where equilibrium bond lengths were relatively insensitive to the regularization strengths.\cite{lee_two_2019}
The recommended regularization values from molecular benchmarks (0.7 for $\sigma$-MP2, \cite{shee_regularized_2021} 1.1 for $\kappa$-MP2, \cite{shee_regularized_2021, rettig_revisiting_2022} 1.45 for $\kappa$-OOMP2 \cite{lee_regularized_2018}) give good lattice constants compared to experiments, but significantly underestimate the cohesive energies. A much higher regularizer value needs to be used to get close to experimental cohesive energy, but comes at the trade-off of severely underestimating the equilibrium lattice constant. This suggests that an alternative regularization strategy is necessary to handle metallic systems.

We now turn to examining the performance of BW-s2. BW-s2 with $\alpha = 1$ gives a cohesive energy of 1.67 eV/atom, which is only 0.01 eV per atom (0.23 kcal/mol) away from the experiment. \cite{schimka_improved_2011} 
At the same time, the equilibrium lattice constant is predicted to be 3.44~{\AA} close to the experimental value (3.45~{\AA}). 
While this is already exceptionally accurate, one could tune the regularization strength for further improvement.
Molecular benchmark studies have shown that BW-s2($\alpha = 4$) performs well for general chemical applications.~\cite{carter-fenk_optimizing_2023, footnote1} However, in \cref{fig:1}, 
we observe that an $\alpha$ value between 1 and 2 provides a good balance for describing both cohesive energy and equilibrium lattice constant.  
We choose $\alpha = 2$ as a sufficiently strong regularization strength to use for the remainder of this work, considering the transferability to molecular contexts~\cite{carter-fenk_optimizing_2023} and other solid-state systems.

How well do these regularized methods compare to other many-body methods?
In \cref{fig:2}, we compare our methods against the available methods in the literature from three main categories: hybrid DFT, coupled cluster (CC), and quantum Monte Carlo (QMC). First, one of the most accurate DFT functionals for properties of BCC Li is HSE06,\cite{heyd_hybrid_2003, heyd_erratum_2006} which produces accurate cohesive energy, equilibrium lattice constant, and bulk modulus.\cite{zhang_performance_2018} 
We see that dRPA@HF significantly underestimates the cohesive energy and overestimates the bulk modulus, consistent with previous benchmarks on the performance of dRPA for metallic solids. \cite{harl_assessing_2010, ye_periodic_2024} 
For CC methods, \cite{neufeld_ground-state_2022, PhysRevLett.131.186402, ye_periodic_2024} the most expensive CCSDT performs well, while CCSD and modified CCSD(T) (CCSD(T)$_{SR}$) both have relatively large errors greater than 0.2 eV for the cohesive energy. 
Finally, variational QMC (VMC) gives accurate cohesive energy estimates but overestimates the equilibrium lattice constant significantly.\cite{PhysRevB.54.8393} 
Both $\kappa$-MP2 and $\sigma$-MP2 fail to predict accurate energies and properties of Li.
Compared to all these methods, we find an $O(N^5)$ method, BW-s2, outperforms an $O(N^6)$ method, CCSD, and rivals an $O(N^8)$ method, CCSDT, for cohesive energy and equilibrium lattice constant, and bulk modulus.

\begin{figure*}[hbt!]
    \centering
    \includegraphics[width=\linewidth]{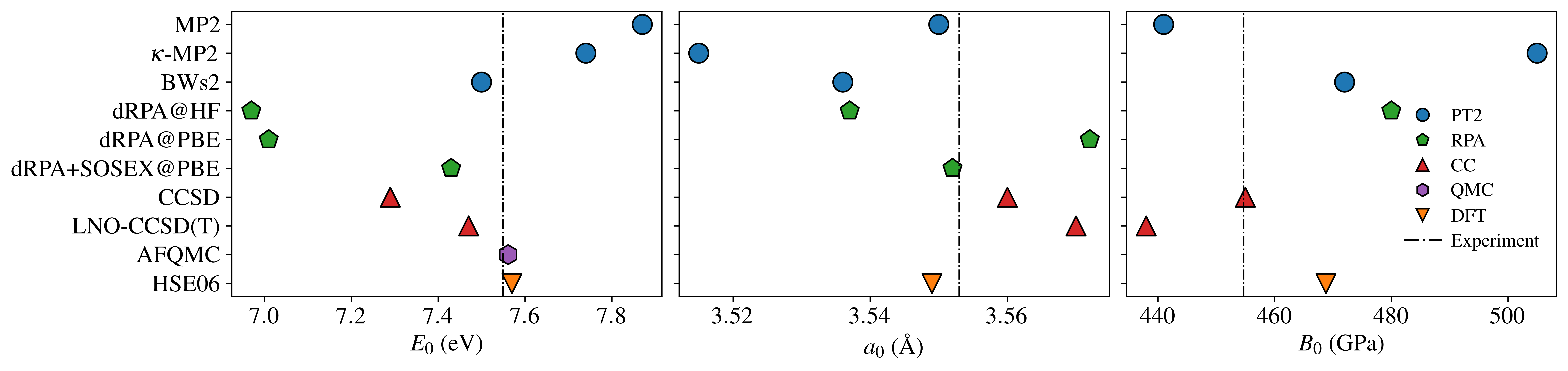}
    \caption{Cohesive energy ($E_0$), equilibrium lattice constant ($a_0$), and bulk modulus ($B_0$) of diamond calculated with different methods compared with experimental values. $\kappa = 1.1$ is used for $\kappa$-MP2, and $\alpha = 2.0$ for BW-s2. dRPA@PBE and dRPA+SOSEX@PBE results are from Ref. [\hspace{-5pt} \citenum{harl_accurate_2009}]. Coupled cluster (CC) results are from Ref. [\hspace{-5pt} \citenum{ye_periodic_2024}]. Auxiliary-field quantum Monte Carlo (AFQMC) result is from Ref. [\hspace{-5pt} \citenum{malone_accelerating_2020}]. HSE06 result is from Ref. [\hspace{-5pt} \citenum{zhang_performance_2018}]. Experimental data is taken from Ref. [\hspace{-5pt} \citenum{schimka_improved_2011}], and includes ZPE correction.
    }
    \label{fig:3}
\end{figure*}

{\it Diamond.}
One natural follow-up investigation is assessing the utility of these regularized methods for simple semiconductors. We chose diamond, as it is a good model for typical wide gap semiconductors. Experimental cohesive energy, equilibrium lattice constant, and bulk modulus (with ZPE correction) are 7.55 eV, 3.55 $\mathrm{\AA}$, and 455 GPa respectively.~\cite{schimka_improved_2011} For this system, both dRPA@HF and dRPA@PBE underestimate the binding energy, which matches the trends for covalent crystals known from literature. \cite{harl_assessing_2010, paier_assessment_2012, ye_periodic_2024} A renormalized version of dRPA that includes SOSEX and rSE corrections partially remedies the underestimation. \cite{paier_assessment_2012, ren_renormalized_2013} On the other hand, MP2 overestimates the binding energy, also well established in literature, \cite{gruneis_second-order_2010, ye_periodic_2024} and one would hope that regularized methods would improve over MP2.

In \cref{fig:3}, the regularized perturbation theory methods ameliorate the over-binding tendencies of MP2. BW-s2($\alpha = 2$) performs better than $\kappa$-MP2, yielding cohesive energy of 7.50 eV per atom, only 0.05 eV away from the experimental reference. Compared to MP2, this presents a significant improvement. At the same time, it slightly underestimates the equilibrium lattice constant and overestimates the bulk modulus. Furthermore, BW-s2 is somewhat better than an existing fifth-order method, RPA+SOSEX@PBE, and a sixth-order method, CCSD, for cohesive energy. It is slightly worse than other state-of-the-art methods such as local natural orbital CCSD(T) (LNO-CCSD(T)),\cite{ye_periodic_2024} auxiliary-field quantum Monte Carlo (AFQMC),\cite{malone_accelerating_2020} and HSE06.\cite{zhang_performance_2018} Overall, while more testing is highly desirable, we conclude that BW-s2 is competitive with state-of-the-art methods for wide bandgap materials such as diamond. As RPA-based $O(N^5)$ methods such as RPA+SOSEX+rSE have not seen much success for metals,\cite{ren_renormalized_2013} these results suggest the unique strength of BW-s2 in simulating bulk metals and semiconductors.

{\it Benzene crystal.}
Another class of materials that we expect MP2 to fail on is systems involving dispersion interactions, such as $\pi$-$\pi$ stacking interactions. Molecular benchmarks have consistently shown that MP2 tends to overestimate the binding energy for these systems, \cite{janowski_benchmark_2012, nguyen_divergence_2020} and regularized perturbation theory has been shown to offer improvements for these molecular systems.\cite{loipersberger_exploring_2021,shee_regularized_2021, carter-fenk_optimizing_2023} 
To examine the efficacy of regularization for dispersion-stabilized solids, we investigated the lattice formation energy of the benzene crystal at its 138K lattice geometry.~\cite{bacon_crystallographic_1997} 
MP2 has been shown to significantly overestimates the cohesive energy of this system \cite{ringer_first_2008, del_ben_second-order_2012, bintrim_integral-direct_2022} compared to the best theoretical estimate, 55.9 $\pm$ 0.86 kJ/mol, and the experimental estimate, 55.3 $\pm$ 2.2 kJ/mol.\cite{yang_ab_2014} 
dRPA+rSE@PBE is an accurate method for this system,\cite{stein_massively_2024} consistent with the well-known strength of dRPA in describing $\pi$-$\pi$ interactions. \cite{lu_ab_2009, lebegue_cohesive_2010}

\begin{figure}[hbt!]
    \centering
    \includegraphics[width=\linewidth]{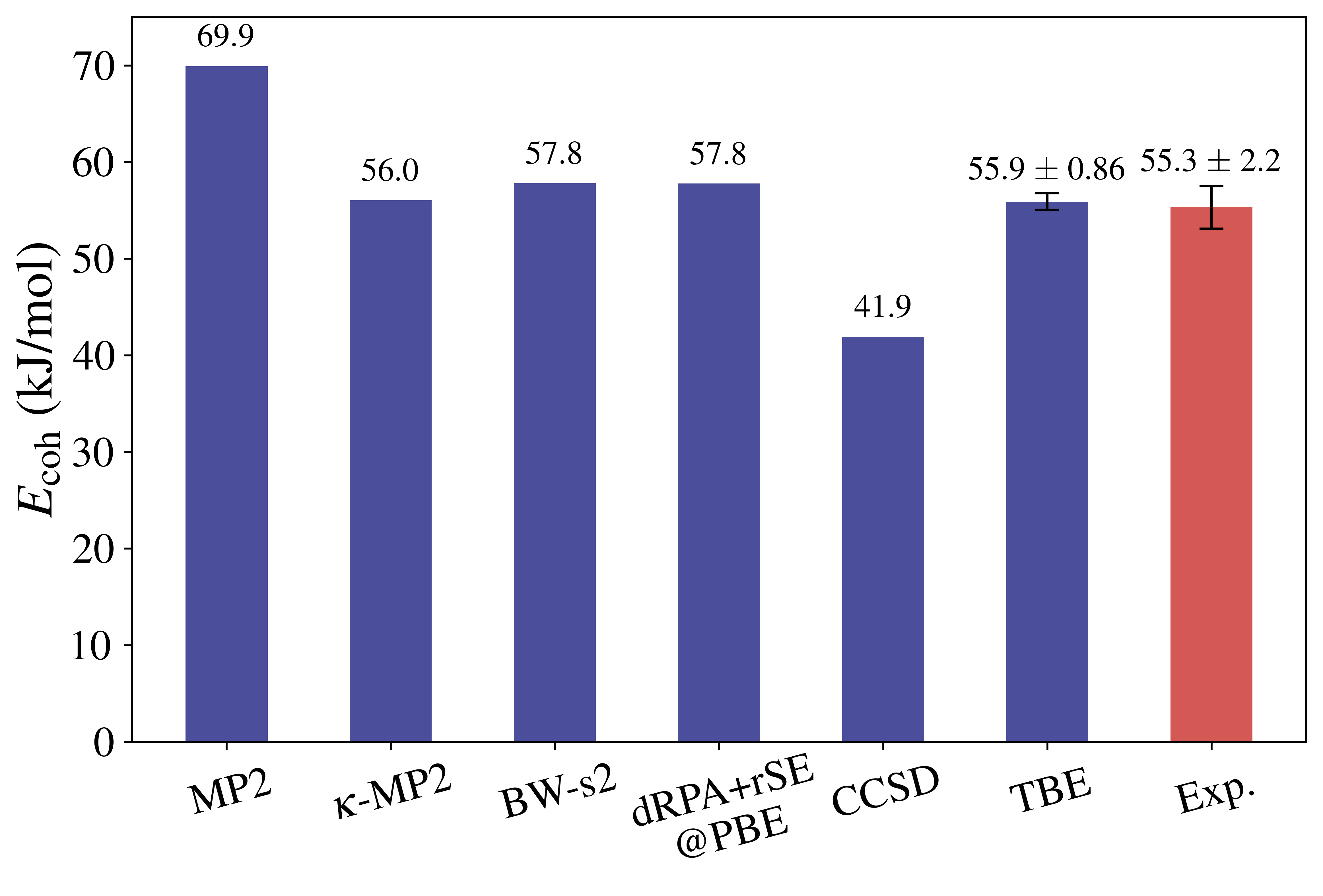}
    \caption{Cohesive energy (eV) of the benzene crystal calculated with different methods. $\kappa = 1.1$ is used for $\kappa$-MP2, $\alpha = 2.0$ for BW-s2. dRPA+rSE@PBE result is from Ref. [\hspace{-5pt} \citenum{stein_massively_2024}]. CCSD and theoretical best estimate (TBE) value is from Ref. [\hspace{-5pt} \citenum{yang_ab_2014}]. Experimental (Exp.) value from Ref. [\hspace{-5pt} \citenum{roux_critically_2008}] and [\hspace{-5pt} \citenum{yang_ab_2014}].
    }
    \label{fig:4}
\end{figure}

\begin{figure*}[hbt!]
    \centering
    \includegraphics[width=\linewidth]{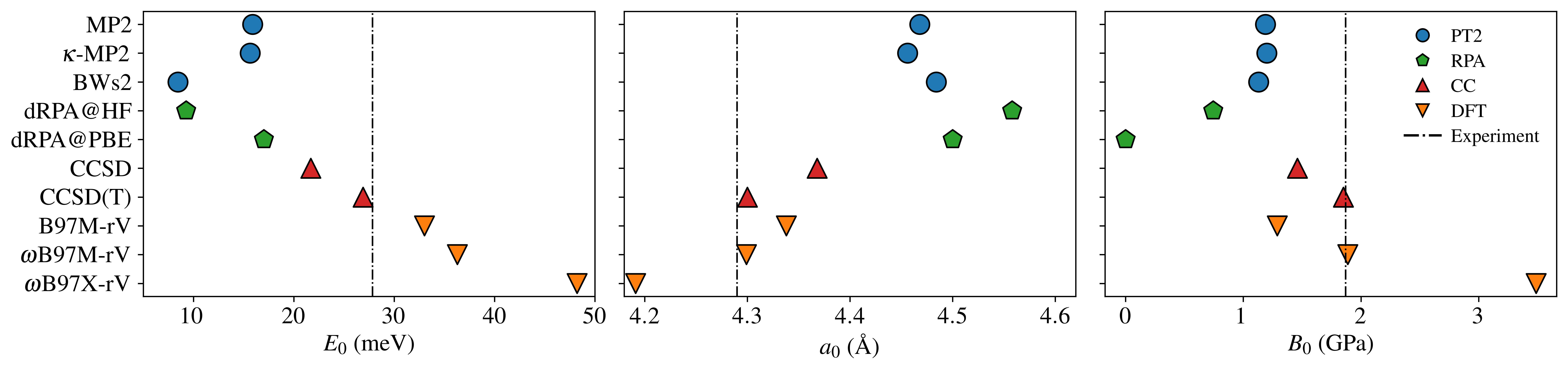}
    \caption{Cohesive energy ($E_0$), equilibrium lattice constant ($a_0$), and bulk modulus ($B_0$) of FCC neon crystal calculated with different methods compared with experimental values. $\kappa = 1.1$ is used for $\kappa$-MP2, and $\alpha = 0.5$ for BW-s2. dRPA@PBE result is from Ref. [\hspace{-5pt} \citenum{harl_cohesive_2008}]. Coupled cluster (CC) and experimental results (including ZPE corrections from CCSD(T)) are from Ref. [\hspace{-5pt} \citenum{schwerdtfeger_convergence_2010}].}
    \label{fig:5}
\end{figure*}

In \cref{fig:4}, we present the cohesive energy for the benzene crystal calculated with MP2, $\kappa$-MP2, BW-s2, and dRPA+rSE, compared to other theoretical and experimental values found in the literature. The best theoretical estimate is based on local CCSD(T) with composite corrections. MP2 significantly overestimates the cohesive energy by about 30\%. 
Both $\kappa$-MP2 and BW-s2 more or less produce
exact cohesive energies, within the uncertainty from the experimental value.
This accuracy rivals the state-of-the-art quartic-scaling dRPA+rSE method.
Remarkably, both perform much better than CCSD.
This example reassures that the regularization parameter of $\alpha = 2$ for BW-s2 may work well for metals, semiconductors, and dispersion-stabilized crystals.

{\it FCC Neon.}
Our last example is the face-centered cubic (FCC) crystal of neon as a representative system for rare gas crystals.~\cite{schwerdtfeger_convergence_2010} 
Unlike most dispersion interactions, this is an example where MP2 underestimates the total lattice energy. 
Hence, any regularization would result in worse performance, serving as a stress test for regularized methods.
The best answer for regularized methods is to recover MP2 results via negligible regularization.
Another important note on this system is that modern dispersion-corrected density functionals are known to struggle for rare gas crystals. \cite{yousaf_applications_2010, ajala_assessment_2019}
For comparison, we will present results obtained by combinatorially optimized density functionals containing the VV10 dispersion correction, \cite{vydrov_nonlocal_2010}  B97M-rV,\cite{mardirossian_mapping_2015, mardirossian_use_2017} $\omega$B97X-rV,\cite{mardirossian_b97x-v_2014}, and $\omega$B97M-rV. \cite{mardirossian_b97m-v_2016} 
CCSD(T) is known to be the most accurate method for this system, with the zero-point energy (ZPE) corrected cohesive energy reaching less than 5\% error of the experimental value.\cite{schwerdtfeger_convergence_2010} 
Experimental results are 4.464~{\AA} equilbrium lattice constant, 
27 meV lattice energy, 
and 1.9 GPa bulk modulus.\cite{batchelder_measurements_1967}

In \cref{fig:5}, it can be seen that all the perturbation-theory-based methods underestimate the cohesive energy compared to the experimental and CCSD(T) results. The tendency of MP2 to underestimate the cohesive energy for neon solid crystal has been reported before, \cite{schwerdtfeger_convergence_2010} and it was found that the MPn series beyond MP2 mitigates this issue by including higher-order many-body effects. 
While $\kappa$-MP2 minimally changes the MP2 results,
we found that BW-s2 underestimates the cohesive energy more severely compared to MP2, with the recommended $\alpha = 2$ even predicting non-binding results. A weaker regularization strength of $\alpha = 0.5$ predicts the binding but still severly underestimates the cohesive energy compared to MP2, as shown in \cref{fig:5}.
The negligible effect of regularization in $\kappa$-MP2 can be related to the large band gap ($>$ 25 eV at HF level) in the neon crystal.
dRPA@HF is similar to BW-s2 and dRPA@PBE is similar to MP2, suggesting the lack of many-body dispersion in dRPA methods.
The inclusion of (T) on top of CCSD is essential, as demonstrated by the error of CCSD. 
The DFT results seem quite sensitive to the choice of functionals, even though we are testing the B97-family functionals.
Nonetheless, B97M-rV and $\omega$B97M-rV perform comparably to CCSD, presenting opportunities for the design of double-hybrid functionals based on B97-family functionals using the regularized correlation form.

{\it Conclusions.} In summary, among the regularized methods considered in this work, BW-s2 was found to perform well for metallic, wide band-gap, and dispersion-stabilized solids. This is the most robust $O(N^5)$ method, often surpassing the accuracy of CCSD. Nevertheless, for neon crystals, BW-s2 and $\kappa$-MP2 do not offer any improvement over MP2, leaving room for further improvement.
Looking forward, we are interested in further exploring different physically-motivated quasi-particle energy correction schemes \cite{coveney_regularized_2023, dittmer_repartitioning_2025} and their effect on regularizing the second- and third-order energies.~\cite{loipersberger_exploring_2021}
Incorporation of these regularized perturbation theory methods as part of double-hybrid functionals~\cite{goerigk2014double} is also a direction worth exploring in the search for generalizable, accurate, and affordable theoretical methods for calculations on large-scale solid-state systems. Lastly, charged and charge-neutral excitations are vastly unexplored with these regularized methods. We expect significantly improved band structures and exciton properties compared to MP2.\cite{gruneis_second-order_2010, iskakov_effect_2019} 

{\it Acknowledgements.} This work was supported by Harvard University’s startup funds. This work used computational resources from FASRC cluster supported by the FAS Division of Science Research Computing Group at Harvard, and the Delta system at the National Center for Supercomputing Applications as well as the Anvil system at Purdue University through allocation CHE250065 from the Advanced Cyberinfrastructure Coordination Ecosystem: Services \& Support (ACCESS) program, supported by National Science Foundation grants \#2138259, \#2138286, \#2138307, \#2137603, and \#2138296. We thank Tim Berkelbach, Kay Carter-Fenk, and Martin Head-Gordon for useful discussions. Data for this work can be found at Ref. [\hspace{-7pt}\citenum{data}].

\newpage
\appendix
\makeatletter
\let\rvtx@orig@subsubsection\subsubsection
\renewcommand{\subsubsection}[1]{%
  \vspace{-0.8\baselineskip}%
  \rvtx@orig@subsubsection{#1}%
}
\makeatother
\renewcommand{\theequation}{A\arabic{equation}}
\setcounter{equation}{0}
\renewcommand{\thefigure}{A\arabic{figure}}
\setcounter{figure}{0}

\section*{Appendix}
Throughout this work, we use $\mu, \nu, ...$ to represent atomic orbital (AO) indices, $i, j, ...$ to represent occupied molecular orbital (MO) indices, and $a, b, ...$ to represent virtual MO indices. Unless otherwise noted, spin-restricted HF orbitals are used as the reference for our correlation energy calculations.
For our periodic calculations, we use a crystalline GTO basis of the form:
\begin{equation}
\phi_{\mu_{\mathbf k}}(\mathbf r) = \frac{1}{\sqrt{N_k}} \sum_{\mathbf R} \exp(i\mathbf k \cdot \mathbf R) \phi_\mu(\mathbf r - \mathbf R).
\end{equation}    
Where $\mathbf{R}$ is the real space lattice vectors, $\mathbf{k}$ is crystalline momentum, $\phi_\mu(\mathbf{r})$ is the $\mu$-th contracted Gaussian type orbital (CGTO) basis function, and $N_k$ is the number of $\mathbf k$-points sampled.
Crystalline GTOs in this form are normalized over the entire Born-von Karman supercell.
\subsection{Derivation of periodic BW-s2}\label{app:A}
Here, we present the derivation of periodic BW-s2, following closely the framework laid out in Refs. [\hspace{-5pt} \citenum{dittmer_repartitioning_2025}] and [\hspace{-5pt} \citenum{carter-fenk_repartitioned_2023}].
We start by writing the standard second-order Brillouin-Wigner perturbation theory expression:\cite{brillouin_problemes_1932, ostlund_perturbation_1975}
\begin{align}
    E^{(2)}_{0} = \sum_{m} \frac{|\bra{\psi_0^{(0)}}\hat{H}_1\ket{\psi_m^{(0)}}|^2}{E^{(0)}_0 + E^{(1)}_0 + E^{(2)}_0 - E^{(0)}_m}
\end{align}
If we insert the conventional Møller-Plesset partitioning of the Hamiltonian where $\hat{H}_0 = \hat{F}$ and $\hat{H}_1 = \hat{H} - \hat{F}$ and apply the Slater-Condon rules, we obtain:
\begin{align} \label{eq:BW2}
    E^{(2)}_{0} = -\frac{1}{4}\sum_{ijab} \frac{|\bra{ij}\ket{ab}|^2}{\Delta_{ij}^{ab} - E^{(1)}_0 - E^{(2)}_0}
\end{align}
where $\Delta_{ij}^{ab} = \varepsilon_a + \varepsilon_b - \varepsilon_i - \varepsilon_j $. This expression is clearly not size-consistent since $E^{(1)}$ and $E^{(2)}$ are size-extensive quantities. 

The idea of BW-s2 is using a different partitioning of the Hamiltonian to cancel out the global energy terms in the denominator, $E^{(1)}$ and $E^{(2)}$. 
To achieve this goal, we rewrite $\hat{H}_0$ in second quantization as:
\begin{align}
    &\hat{H}_0 = \sum_{pq} (F_{pq} + W_{pq} + C\delta_{pq}) \hat{a}^\dagger_p \hat{a}_q \\
    &\hat{H}_1 = \hat{H} - \hat{H}_0,
\end{align}
where $\mathbf W$ is a regularizer and $C$ is some constant.
The choice of how to re-partition the Hamiltonian while retaining size-consistency is not unique, as discussed extensively in Ref [\hspace{-5pt} \citenum{dittmer_repartitioning_2025}]. 
In BW-s2, we make a specific choice for $\mathbf{W}$,~\cite{carter-fenk_repartitioned_2023}
\begin{align}
    W_{ij} = \frac{1}{8} \sum_{kab} \left[ t_{ik}^{ab}\langle jk||ab \rangle + t_{jk}^{ab}\langle ik||ab \rangle \right]
\end{align}
and $W_{ia} = W_{ab} = 0$. We also observe $\mathrm{Tr}(\mathbf W) = E^{(2)}$. 
Along with $C = E^{(1)}$, this leads to the new $E^{(0)^{'}}_i = F_{ii} + W_{ii} + C$, and $E^{(1)^{'}}_i = E^{(1)} - F_{ii} - W_{ii} - C$. Plugging this expression back into \cref{eq:BW2}, the global energy terms in the denominator are canceled out, making the overall energy expression size-consistent.

For periodic systems, due to momentum conservation, the Fock matrix and any other momentum-conserving one-electron operators are block-diagonal in $\mathbf{k}$. Similarly, $\mathbf{W}$, in this case, is also block-diagonal in $\mathbf{k}$, and it is given as
\begin{align}
    W_{ij}^{\mathbf{k}_1} &= \frac{1}{8} \sum_{kab} \sum_{\mathbf{k}_2 \mathbf{q}} \left[ t_{i_{\mathbf{k}_1}k_{\mathbf{k}_2+\mathbf{q}}}^{a_{\mathbf{k}_1+\mathbf{q}}b_{\mathbf{k}_2}} \langle j_{\mathbf{k}_1}k_{\mathbf{k}_2+\mathbf{q}}||a_{\mathbf{k}_1+\mathbf{q}}b_{\mathbf{k}_2}\rangle \right. \\ \nonumber
    & \hspace{20pt} + \left. t_{j_{\mathbf{k}_1}k_{\mathbf{k}_2+\mathbf{q}}}^{a_{\mathbf{k}_1+\mathbf{q}}b_{\mathbf{k}_2}} \langle i_{\mathbf{k}_1}k_{\mathbf{k}_2+\mathbf{q}}||a_{\mathbf{k}_1+\mathbf{q}}b_{\mathbf{k}_2}\rangle \right].
\end{align}
Using this, we can obtain the modified occupied orbital energies via \cref{eq:dressed}.

\subsection{Computational details}\label{app:B}
All calculations were carried out with a development version of Q-Chem.\cite{epifanovsky_software_2021} We used the Goedecker-Teter-Hutter (GTH) pseudopotential optimized for Hartree-Fock (GTH-HF-rev) \cite{goedecker_separable_1996, hutter_new_2018} and the correlation-consistent GTH-cc-pVXZ Gaussian basis set \cite{ye_correlation-consistent_2022} (GTH-aug-cc-pVXZ for noble gases) for all calculations. 
The Gaussian-Planewave (GPW) density fitting algorithm \cite{goedecker_separable_1996, vandevondele_quickstep_2005} was used for evaluating the two-electron integrals. 
The Madelung correction\cite{fraser_finite-size_1996} was employed to treat the divergence of exact exchange in crystalline systems.
This correction was included in the orbital energies of the subsequent correlation calculations.
All k-point calculations were performed using a Monkhorst-Pack grid \cite{monkhorst_special_1976} shifted to contain the Gamma point. Counterpoise correction \cite{boys_calculation_1970} was used in all of our cohesive energy calculations. 

Total energies of solids were extrapolated to the thermodynamic limit (TDL) via:
\begin{equation}
    E_{\mathrm{TDL}} = \frac{E_{N_{k_1}}N_{k_1}^{-1} - E_{N_{k_2}}N_{k_2}^{-1}}{N_{k_1}^{-1} - N_{k_2}^{-1}}
\end{equation}
which assumes the linear behavior in $1/N_k$ with Madelung correction. The mean field energies are extrapolated to the complete basis set (CBS) limit by an exponential form, \cite{halkier_basis-set_1999} and the correlation energies are extrapolated using the two point formula:  
\begin{equation}
    E_{\mathrm{CBS}} = \frac{E_{X_1}X_1^{3} - E_{X_2}X_2^{3}}{X_1^{3} - X_2^{3}}
\end{equation}
where $X$ is the cardinal number of the basis set, which assumes the linear behavior in $n_{vir}^{-1}$ for correlation energies.\cite{shepherd_convergence_2012}  

\subsubsection{Cholesky decomposition}
To accelerate the evaluation of correlation energies, we used the Cholesky decomposition of the electron repulsion integrals (ERIs), which gives us a systematic control of the errors arising from the low-rank approximation with a single threshold. 
Cholesky vectors of the ERIs, effectively two-electron, three-center integrals, were obtained using the modified Cholesky decomposition algorithm.\cite{beebe_simplifications_1977, koch_reduced_2003} 
This offers the following factorization:
\begin{align}
(\mu_{\mathbf{k}_1}\nu_{\mathbf{k}_1+\mathbf{q}} |
\lambda_{\mathbf{k}_2+\mathbf{q}} \sigma_{\mathbf{k}_2}) = \sum_P^{N_\mathrm{Chol}}L^{P, \mathbf{q}}_{\mathbf{k}_1,\mu,\nu}
(L^{P, \mathbf{q}}_{\mathbf{k}_2,\sigma,\lambda})^*
,
\end{align}
where the number of Cholesky vectors, $N_\mathrm{Chol}$, is controlled by the convergence threshold of the Cholesky decomposition algorithm.

\subsubsection{Integral-direct implementation via THC}
For larger systems, the modified Cholesky decomposition algorithm that must be run for each $\mathbf{q}$ point might become prohibitively expensive, and the storage cost of the Cholesky vectors also becomes another immediate concern. 
For these cases, we used tensor hypercontraction (THC), also known as interpolative separable density fitting (ISDF), \cite{hohenstein_tensor_2012,lu_compression_2015, lee_systematically_2020, rettig_even_2023, yeh_low-scaling_2023} to further compress the storage requirements and reconstructed Cholesky vectors on-the-fly from the THC vectors. 

In the THC format, decomposition of the ERIs is given by:
\begin{align} \label{eq:thc}
&(\mu_{\mathbf{k}_1}\nu_{\mathbf{k}_1+\mathbf{q}} |
\lambda_{\mathbf{k}_2+\mathbf{q}} \sigma_{\mathbf{k}_2}) \\ \nonumber
& = \sum_{\hat P\hat Q}^{N_{\mathrm{ISDF}}} 
     (X_{\mu_{\mathbf{k}_1}}^{\hat P})^*
     X_{\nu_{\mathbf{k}_1+\mathbf{q}}}^{\hat P} 
     M_{{\hat P}{\hat Q}}^{\mathbf{q}}
     (X_{\lambda_{\mathbf{k}_2+\mathbf{q}}}^{\hat Q})^*
     X_{\sigma_{\mathbf{k}_2}}^{\hat Q},
\end{align}
where $\hat{P}$ is the set of interpolating points chosen via the Cholesky-based algorithm \cite{yeh_low-scaling_2023, matthews_improved_2020} at $\mathbf{q} = 0$. 
We then determine the THC vectors $\mathbf X$ via a least-squares fit to accurately approximate \cref{eq:thc}.
Similarly to the Cholesky decomposition of ERIs, $N_{\mathrm{ISDF}}$ is controlled by adjusting a single convergence threshold.
Two-electron, three-center integrals, $\mathbf{L}$, can then be reconstructed by:
\begin{align}
    L^{P,\mathbf{q}}_{\mathbf{k}_1,\mu,\nu} = \sum_{\hat P} 
     (X_{\mu_{\mathbf{k}_1}}^{\hat P})^*
     X_{\nu_{\mathbf{k}_1+\mathbf{q}}}^{\hat P} 
     D_{\hat P P}^{\mathbf{q}}
\end{align}
where $\mathbf{D}$ is the full-rank Cholesky decomposition of $\mathbf{M}$ ($\mathbf{M} = \mathbf{D}\mathbf{D}^\dagger$).

\begin{figure}[hbt!]
    \centering
    \includegraphics[width=0.9\linewidth]{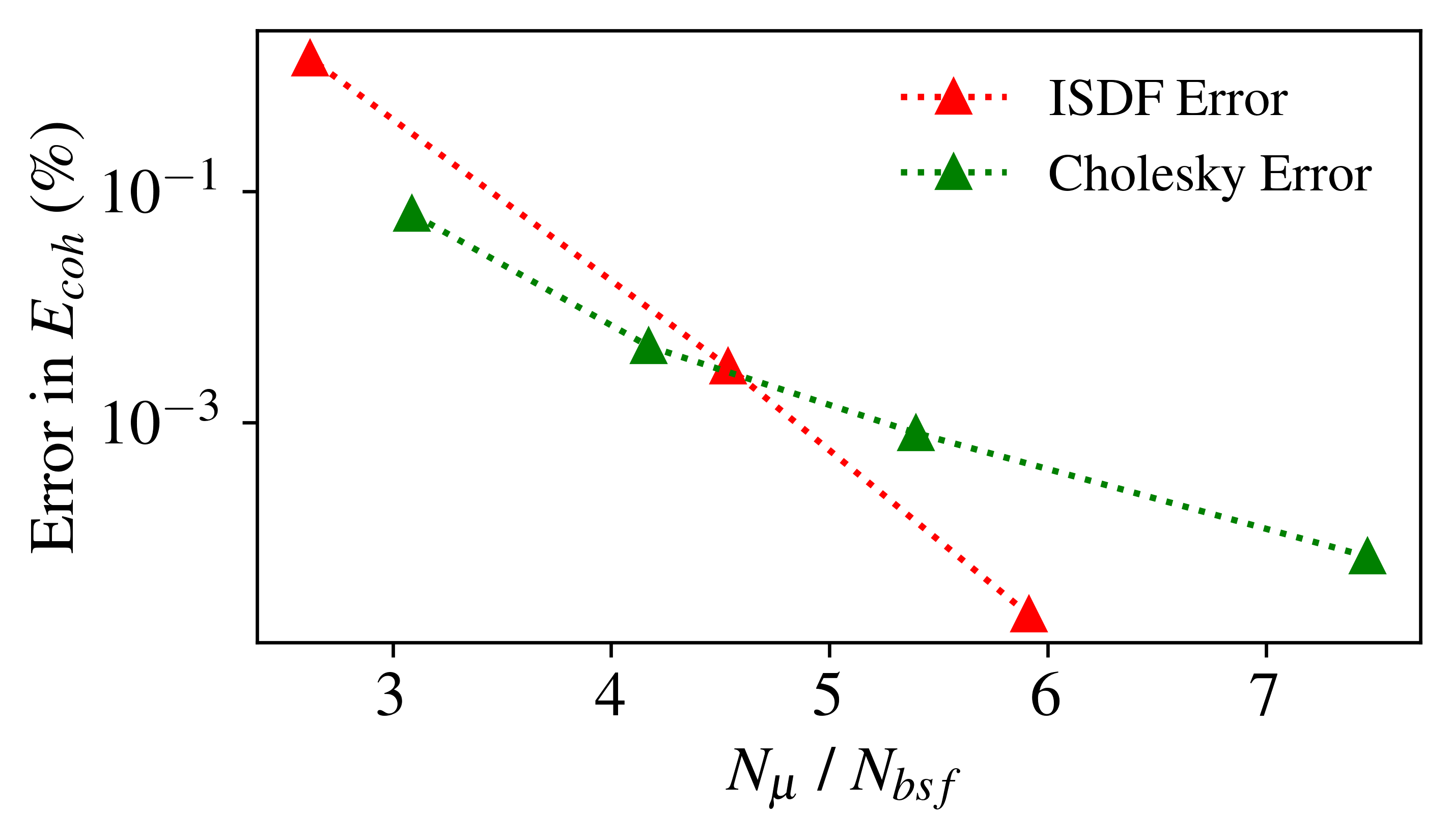}
    \caption{Relative error in the MP2 cohesive energy of diamond plotted against auxiliary basis size ($N_\mu$ = $N_{\mathrm{Chol}}$ or $N_{\mathrm{ISDF}}$).
    } 
    \label{fig:sup1}
\end{figure} 

For our methods, we only need the occupied-virtual block of the ERI tensor. Hence, we perform the least-squares fit directly to this block, resulting into a more compact THC compression.
In \cref{fig:sup1}, we show the convergence of the MP2 cohesive energy of diamond with respect to the auxiliary basis size. It can be seen that the relative error in the cohesive energy can be efficiently reduced to under $10^{-3}$ \% using an auxiliary basis size of around $N_{\mathrm{Chol}} / N_{\mathrm{ISDF}}\approx 5 \times N_{\mathrm{bsf}}$. 

\subsubsection{Details on BCC Li}
For Li, we set the kinetic energy cutoff for the auxillary plane-wave basis set used in GPW to be 1000 eV. 
The lattice energies were calculated using up to $5^3$ $\mathbf k$-mesh and GTH-cc-pVQZ basis set, extrapolated to the TDL using $4^3$ and $5^3$ $\mathbf k$-meshes, and  to the complete basis set (CBS) limit using the T-Q extrapolation with a $1/X^3$ form, where $X$ is the cardinal number of the basis set. 
The calculated cohesive energies are then fitted using the Birch-Murnaghan equation of state \cite{birch_finite_1947, murnaghan_compressibility_1944} to obtain the equilibrium lattice constant and bulk modulus.

\subsubsection{Details on diamond}
For diamond, we use 500 eV as the kinetic energy cutoff for the auxillary plane-wave basis set used in GPW. 
The diamond lattice energies were extrapolated to the TDL using $4^3$ and $5^3$ results, and to the CBS limit using T-Q extrapolation. 
The equilibrium properties obtained from the Birch-Murnaghan fit are compared to results from different theoretical and experimental studies found in the literature, as shown in \cref{fig:5}. 

\subsubsection{Details on benzene crystal}
\begin{figure}[hbt!]
    \centering
    \includegraphics[width=\linewidth]{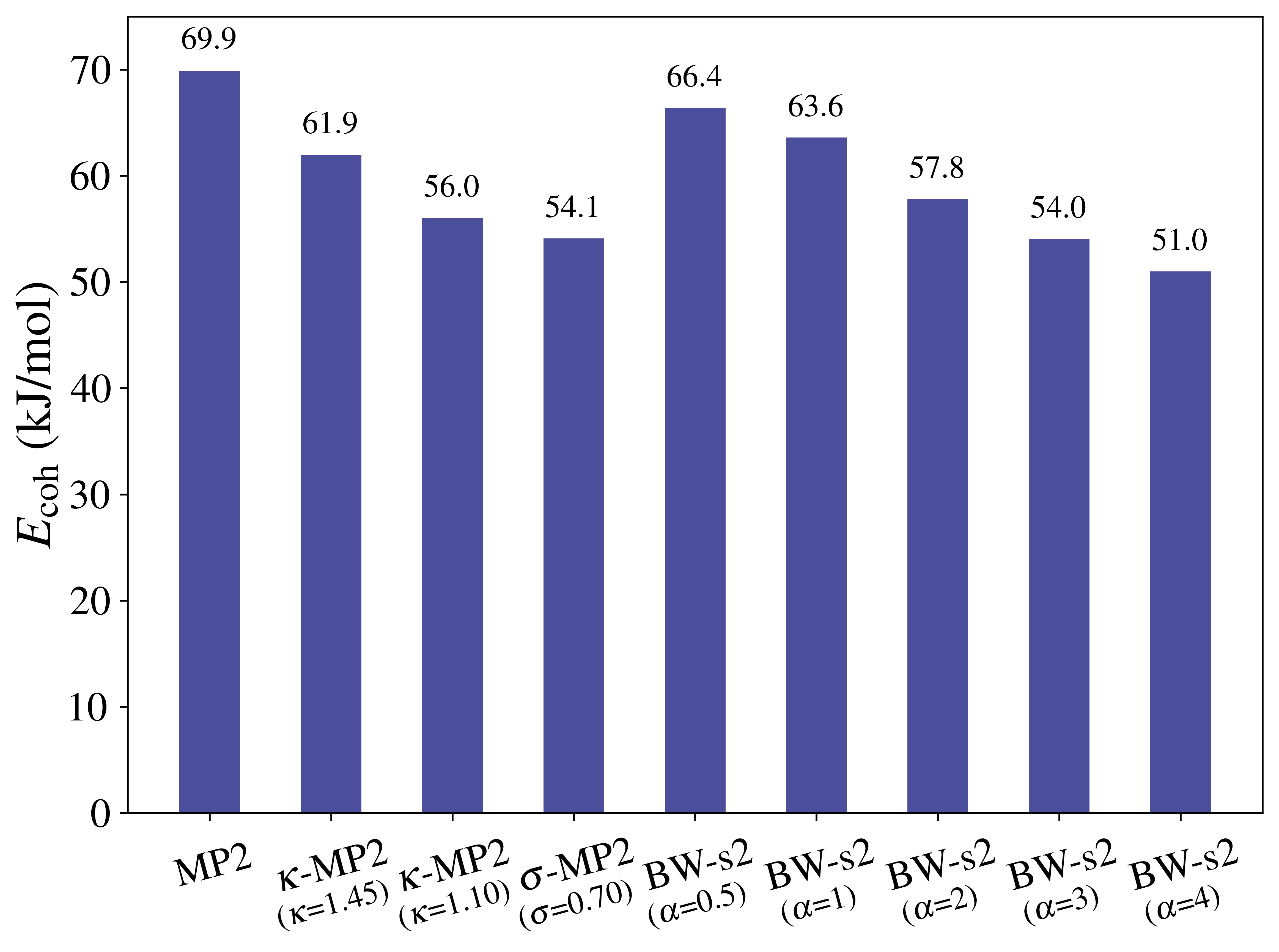}
    \caption{Cohesive energy of benzene crystal calculated with different regularization strengths for $\kappa$-MP2, $\sigma$-MP2, and BW-s2($\alpha$).
    } 
    \label{fig:sup2}
\end{figure}

For this system, we use 750 eV as the kinetic energy cutoff for the auxiliary plane-wave basis set used in GPW. 
Previous studies have shown that careful TDL/CBS extrapolation is essential for obtaining precise cohesive energies for this system.\cite{bintrim_integral-direct_2022} 
For our results, we extrapolate the energies of the benzene crystal to the TDL using $2^3$ and $3^3$ $\mathbf{k}$-mesh results and the GTH-cc-pVDZ basis set, and apply the CBS correction obtained from MP2 crystal energies at $2^3$ k-point mesh using GTH-cc-pVTZ and GTH-cc-pVQZ basis. We note that the difference between TDL extrapolation using $2^3$ and $3^3$ $\mathbf{k}$-mesh and using $1^3$ and $2^3$ $\mathbf{k}$-mesh is negligible (around 0.3 kJ/mol).  Our MP2 result is within 3 kJ/mol difference compared to  Ref.~\citenum{bintrim_integral-direct_2022}, likely due to the density fitting errors in their work. In \cref{fig:sup2}, we show an expanded version of \cref{fig:4} that includes the cohesive energy of benzene crystal calculated with different regularization strengths ($\kappa$, $\sigma$, $\alpha$) for the regularized perturbation theory methods. We show the results of $\kappa = 1.1$ and $\alpha = 2$ in \cref{fig:4}.

\subsubsection{Details on FCC Ne}
The cohesive energy of noble-gas crystals is minuscule in magnitude and, therefore, requires a carefully converged calculation setup to obtain accurate energies. 
We found that a kinetic energy cutoff of 3000 eV for the mean field calculation and 2000 eV for the correlation energy calculation was sufficient to reduce the plane-wave density fitting error to less than $10^{-6}$ hartree. HF calculations were extrapolated to the TDL using $5^3$ and $6^3$ $\mathbf k$-mesh results, and checked for linearity using calculations at $8^3$ $\mathbf k$-mesh. The correlation energy calculations were extrapolated using $4^3$ and $5^3$ $\mathbf k$-mesh results. Furthermore, we observed a non-negligible basis set incompleteness error from even the GTH-aug-cc-pVQZ basis set for neon.\cite{ye_correlation-consistent_2022} Hence, we made a custom `GTH-aug-cc-pV5Z' basis set by adding to GTH-aug-cc-pVQZ two primitive GTOs by uncontracting the most diffuse basis functions of contracted 2s and 2p functions in GTH-aug-cc-pVQZ and taking the d, f, g, and h orbitals of the original Dunninng aug-cc-pV5Z basis set.\cite{dunning_gaussian_1989} CBS energies were then obtained using a QZ-5Z extrapolation.

\begin{figure}[hbt!]
    \centering
    \includegraphics[width=\linewidth]{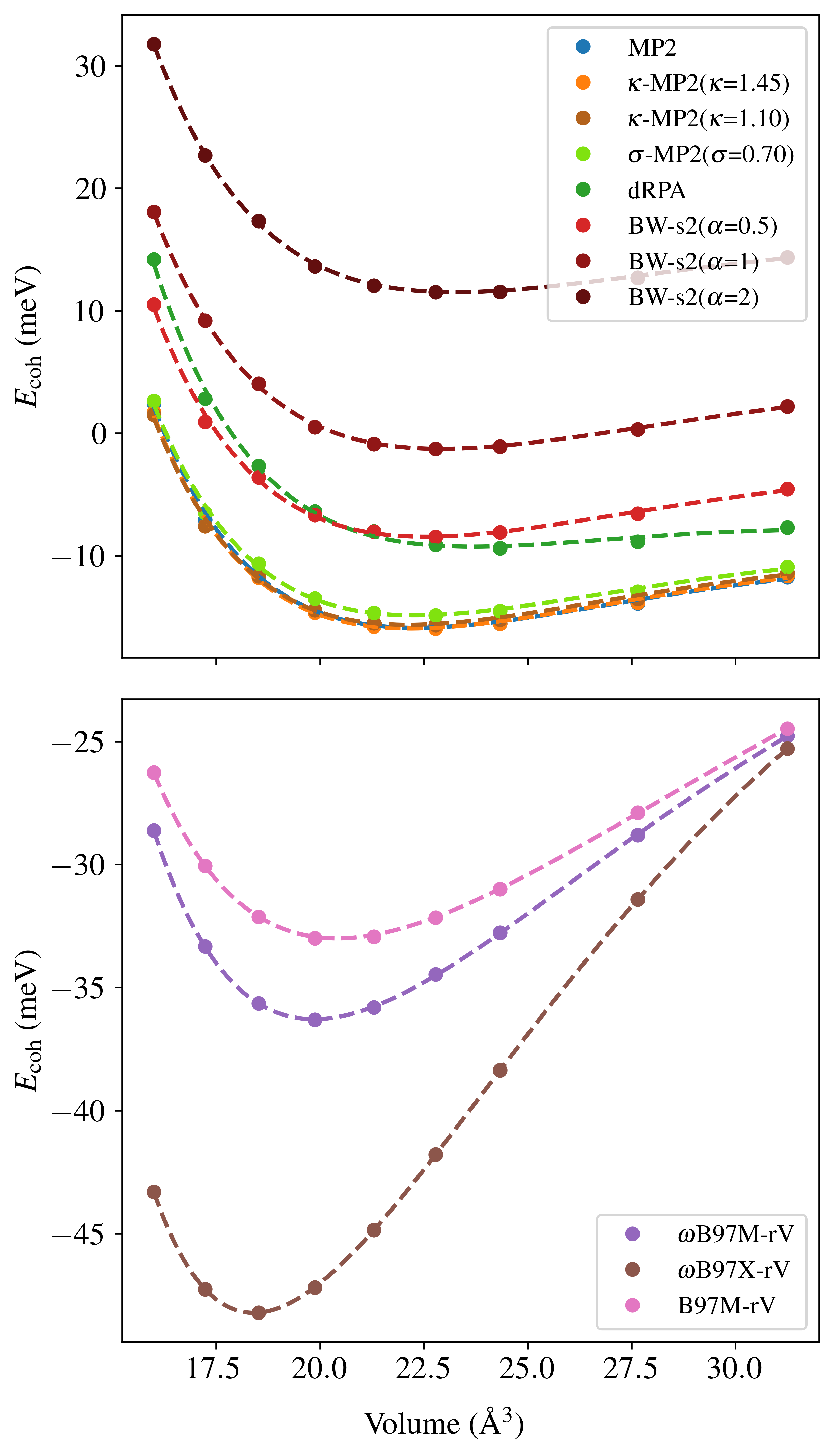}
    \caption{Cohesive energy of neon crystal at lattice constants near equilibrium geometry calculated with different methods. The dashed line represents the Birch-Murnaghan equation-of-state fit.
    } 
    \label{fig:sup3}
\end{figure}

In \cref{fig:sup3}, we show the cohesive energy of neon crystal around the equilibrium lattice constant calculated with wavefunction-based methods and dispersion corrected density functional theory, with the Birch-Murnaghan equation-of-state fit shown as a dashed line. Among the different regularization strengths, we show the cohesive energy, equilibrium lattice constant, and bulk modulus extracted from $\kappa$ = 1.1 and $\alpha$ = 0.5 in \cref{fig:5}. 

\newpage
\clearpage
\bibliography{references_final}
\end{document}